# Can we close the Bohr-Einstein quantum debate?

**Authors:** Marian Kupczynski[1]

**Affiliations:**

[1] Département de l'Informatique, Université du Québec en Outaouais (UQO), Case postale 1250, succursale Hull, Gatineau, Quebec, J8X 3X 7 , Canada

Correspondence to: marian.kupczynski@uqo.ca

Recent experiments allowed concluding that Bell-type inequalities are indeed violated thus it is important to understand what it means and how can we explain the existence of strong correlations between outcomes of distant measurements. Do we have to announce that: Einstein was wrong, Nature is nonlocal and nonlocal correlations are produced due to the quantum magic and emerge, somehow, from outside space-time? Fortunately such conclusions are unfounded because if supplementary parameters describing measuring instruments are correctly incorporated in a theoretical model then Bell-type inequalities may not be proven .We construct a simple probabilistic model explaining these correlations in a locally causal way. In our model measurement outcomes are neither predetermined nor produced in irreducibly random way. We explain in detail why, contrary to the general belief; an introduction of setting dependent parameters does not restrict experimenters' freedom of choice. Since the violation of Bell-type inequalities does not allow concluding that Nature is nonlocal and that quantum theory is complete thus the Bohr-Einstein quantum debate may not be closed. The continuation of this debate is not only important for a better understanding of Nature but also for various practical applications of quantum phenomena.

## I. INTRODUCTION

Since 1982 significant violations of Bell's [1,2] , Clauser-Horn-Shimony–Holt (CHSH) [3], Clauser-Horne [4] and Eberhard's [5] inequalities have been reported in several excellent experiments [6-14]. After closing remaining loopholes in [12-14] the authors concluded that Nature is not governed by *local realism*. In order to avoid misunderstanding it is important to clarify the meaning of *local realism* and to examine implications of violation of various Bell-type inequalities for the debate held by Einstein [15-17] with Bohr [18,19] concerning the interpretation, locality and completeness of quantum mechanics (QM).

To understand correctly why Bell–type inequalities are violated, in spin polarization correlation experiments (SPCE), we examine probabilistic models and assumptions which were used to prove these inequalities.

Bell-type inequalities may be proven using two different hidden variable models. Both of them assume that: *there are no physical influences travelling faster than light* (locality).

In local realistic hidden variable model LRHV [1, 2] one assumes also *realism* defined as: *any system has pre-existing values for all possible measurements* of the system [12]. This assumption is called more precisely *counterfactual definiteness* (CFD) because *it is assuming the definiteness of the results of measurements which were not performed on a given individual system.* CFD *is in striking disagreement with quantum mechanics and the complementarity principle of quantum observables* [20].

Spin projections are not predetermined attributes of physical systems and are only created in interaction of these systems with measuring instruments thus it is not surprising that LRHV cannot explain correlations observed in SPCE.

In stochastic hidden variable models (SHV) outcomes, instead of being predetermined, are *obtained in irreducibly random way* (IR) [2, 4]. However randomly produced outcomes cannot be correlated strongly enough to reproduce correlations observed in SPCE.

The impossibility of explaining these correlations using CFD or IR is often called *quantum nonlocality* and is considered to be *one of the most remarkable phenomena known to modern science* [21].

Inspired by John Bell who in 1975 said: *if a hidden variable theory …agrees with QM it will be non-local* [13, 22] several researchers started to investigate the possibility of hidden *quantum influences* which may propagate faster than light but which cannot be used for faster- than- light communications [23].

When it was shown that any finite speed hidden superluminal influence must necessarily lead to faster-than -light communication [24] a mystery became complete*: no story in space-time can tell us how nonlocal correlations happen, hence nonlocal, quantum correlations seem to emerge , somehow, from outside of space-time* [25]. In a recent book, author concludes: *there is no possible local explanation for certain correlations produced by quantum physics. Nature is nonlocal and God does play dice* [26].

Since QM, quantum field theory and the standard model in particle physics are consistent with the theory of relativity and locality of Nature [27-29] such "explanation" and a new law of Nature called nonlocal *randomness* are difficult to accept.

This is why some physicists still believe in the existence of residual experimental loopholes. Namely if there is a *super-determinism* then experimentalists , in spite of their beliefs, have no a *free will* enabling them to choose experimental settings or perhaps random number generators used to choose these settings may be synchronized [30,31].

Fortunately there is a simpler and more rational explanation [29] of long range correlations predicted by QM and observed in SPCE.

The paper is organized as follows. In section 2 we present a probabilistic model providing a locally causal explanation of observed correlations. In section 3 we compare our model with LRHV, SHV and Bell 1970 model [2, 32]. In section 4 we discuss probabilistic meaning of LRHV and SHV models and importance of experimental protocols and contexts. In section 5 we try to demystify the notion of entanglement. In section 6 we explain in detail , using Bayes Theorem, why setting - dependent hidden variables describing measuring instruments do not constrain experimenters' *free will* to choose their settings as they wish . In section 7 we give arguments against a subjective interpretation of quantum theory called Qbism. In section 8 we argue that Bohr-Einstein quantum debate is still important and may not be closed. Section 9 contains some conclusions.

## 2. LOCAL EXPLANATION OF CORRELATIONS IN SPCE

First of all let us recall that, in general, QM does not give predictions for single outcomes of quantum measurements but only provides probabilistic predictions to be compared with the statistical spread of outcomes observed in repeated measurements, on identically prepared physical systems, in well-defined experimental contexts [19, 33].

Since sharp directions do not exist QM does not predict, contrary to a general belief, strict correlations or anti-correlations in SPCE [29, 34-40]. Since local causality is consistent with macroscopic phenomena we want to explain these observed imperfect correlations in a locally causal way.

The correlations among distant events depend on how the pairing of them is done. In SPCE Alice and Bob use synchronized time windows and the observed events are successive clicks or absence of a click on detectors. These events are coded by 0 and ±1 and can be considered to be the values of some random variables (A, B) describing these random experiments. In order to check predictions of QM and/or CHSH inequality one carefully post-selects samples keeping only events in which both Alice and Bob registered a click in their synchronized time-windows.

A model we propose describes how row data before the post selection are created. It is based on the assumption that signals produced by a source *keep a partial memory of the common past* when they arrive to distant laboratories. Then remote measuring instruments operate in a locally causal way and outcomes may be correlated more strongly, than it was permitted by SHV and LRHV.

We use a notation consistent with [21]. As we explain later this notation may be sometimes confusing. We assume:

1. A source is sending two correlated signals (correlated *pairs of photons*) towards distant measuring settings. These correlated signals are described at a moment of measurement by some, hidden to experimentalists, correlated supplementary parameters $(\lambda_1, \lambda_2) \in \Lambda_1 \times \Lambda_2$ and P $(\lambda_1, \lambda_2)$.

2. We assume *freedom of choice*: Alice and Bob may choose their experimental settings (*x*, *y*) randomly or in any systematic way. We do not believe that a setting chosen by Alice may influence a distant setting chosen by Bob. If choice of the setting is made by random number generators then a probability of choosing a particular setting is given by P(x, y).

3. The measuring instruments used in the setting (*x*, *y*), <u>as perceived by the incoming signals at a moment</u> of measurement, are described by supplementary parameters $\lambda_x \in \Lambda_x$, $\lambda_y \in \Lambda_y$ and probability distributions $P_x(\lambda_x)$ and $P_y(\lambda_y)$.

4. The outcomes a= 0,±1 and b=0, ±1 are created in a <u>deterministic way</u> as function of local parameters describing measuring instruments and a *given pair* : P(a| x, $\lambda_1$, $\lambda_x$)=0 or 1 and P(b| y, $\lambda_2$, $\lambda_y$)=0 or 1 . Outcomes 0 correspond to the absence of a click. If we want to test only CHSH or check quantum predictions for correlations we limit ourselves to correlated time-windows in which both clicks were registered. In this case possible outcomes are a= ±1 and b= ±1 [39, 40].

Using the points 1-4 we obtain <u>a probability of obtaining outcomes (a ,b) when a the setting (x, y) is used</u> P(a ,b| x, y):

$$P(a,b|x,y) = \sum_{\lambda \in \Lambda_{xy}} P(\lambda|x,y) P(a|x,\lambda_1,\lambda_x) P(b|y,\lambda_2,\lambda_y) \tag{1}$$

where $\lambda=(\lambda_1,\lambda_2,\lambda_x,\lambda_y)$, $P(\lambda|x,y) = P(\lambda_1,\lambda_2) P_x(\lambda_x) P_y(\lambda_y)$ and $\Lambda_{xy} = \Lambda_1 \times \Lambda_2 \times \Lambda_x \times \Lambda_y$ are different for each pair of experimental settings. The sums should be replaced by integrals if the variables λ are not discrete. Formula (1) contains <u>enough free parameters to fit experimental data for any setting.</u>

Please note that P (λ |x, y) is <u>not a conditional probability</u> in the usual sense deduced from some joint probability of <u>all</u> possible (x, y, λ ) which does not need to exist. <u>P (λ |x, y) denotes only a probability distribution of supplementary parameters used in a particular setting (x, y).</u>

Nevertheless if P (λ ,x, y) = P (λ |x, y) P (x, y) then using Bayes Theorem P (x, y | λ) ≠ P(x ,y) [ 21,31] . This inequality is treated as a proof that <u>if λ depends on (x, y) then experimenters'</u> *freedom of choice* <u>is compromised .</u> Of course it <u>is not true because a conditional probability P(A|B) does not mean ,in general, that B is a cause of A.</u> We discuss this subtle point in detail in section 6.

If we describe only the data registered by Alice without taking into consideration any data registered by Bob we obtain P (a |x):

$$P(a|x) = \sum_{\lambda_1,\lambda_2,\lambda_x} P(a|x,\lambda_1,\lambda_x) P(\lambda_1,\lambda_2) P_x(\lambda_x) \tag{2}$$

A similar formula can be written for P (b |y). Expectation functions in our model can be written as:

$$E(A,B|x,y) = \sum_{\lambda \in \Lambda_{xy}} A_x(\lambda_1,\lambda_x) B_y(\lambda_2,\lambda_y) P_x(\lambda_x) P_y(\lambda_y) P(\lambda_1,\lambda_2) \qquad (3)$$

where $\Lambda_{xy} = \Lambda_1 \times \Lambda_2 \times \Lambda_x \times \Lambda_y$ , $A_x(\lambda_1,\lambda_x)$ and $B_y(\lambda_2,\lambda y)$ are equal $0, \pm 1$ .

Since in our contextual locally causal model hidden variables depend explicitly on the choice of settings Bell-type inequalities cannot be proven. The outcomes are neither predetermined nor obtained in irreducibly random way.

By no means have we claimed that SPCE should be described using (1-3) instead of QM. We only wanted to show that intuitive explanation of long range correlations observed in SPCE is possible. Using completely different models Hans de Raedt and Kristel Michielsen [41-43] succeeded, to simulate several quantum experiments including SPCE without questioning the locality of Nature.

## 3. LRHV, SHV AND BELL 1970 MODEL

We may prove Bell-type inequalities if we assume that <u>for all</u> the settings: $\Lambda_{xy} = \Lambda = \Lambda_1 \times \Lambda_2$ , $P(\lambda|x,y) = P(\lambda_1,\lambda_2)$, $P(a|x,\lambda_1,\lambda_x) = P(a|x,\lambda_1)$ and $P(b|y,\lambda_2,\lambda_y) = P(b|y,\lambda_2)$ . Then the formula (1) is replaced by a formula used in SHV:

$$P(a,b|x,y) = \sum_{\lambda \in \Lambda} P(\lambda) P(a|x,\lambda_1) P(b|y,\lambda_2) \qquad (4)$$

We may also prove Bell-type inequalities using the formula (3) with $\Lambda_{xy} = \Lambda = \Lambda_1 \times \Lambda_2$, $A_x(\lambda_1,\lambda_x) = A_x(\lambda_1)$ and $B_y(\lambda_2,\lambda y) = B_y(\lambda_2)$ taking values 0 and $\pm 1$ (or only $\pm 1$):

$$E(A,B|x,y) = \sum_{\lambda \in \Lambda} A_x(\lambda_1) B_y(\lambda_2) P(\lambda_1,\lambda_2) \qquad (5)$$

The formula (5) is a discrete version of the original formula proposed by Bell in 1964 [1]:

$$E(a,b) = \int_\Lambda A(\lambda) B(\lambda) \rho(\lambda) d\lambda \qquad (6)$$

LRHV models (5-6) are using the same unique parameter space and the same joint probability distribution to describe different incompatible random experiments (x, y) what is in conflict with Kolmogorov theory of probability and with experimental protocols used in SPCE.

In 1970 Bell [32] incorporated in his model setting- dependent hidden variables describing measuring devices. He demonstrated that if these parameters are averaged out one may prove CHSH inequalities [3].

Let us reproduce his proof using our model. From (3) after a summation in (3) over ($\lambda_x$, $\lambda_y$) we obtain:

$$E(A,B|x,y) = \sum_{\lambda_1,\lambda_2} \bar{A}_x(\lambda_1)\bar{B}_y(\lambda_2) P(\lambda_1,\lambda_2) \qquad (7)$$

where $\bar{A}_x(\lambda_1) = \sum_{\lambda_x} A_x(\lambda_1,\lambda_x) P_x(\lambda_x)$ and $\bar{B}_y(\lambda_2) = \sum_{\lambda_y} B_y(\lambda_1,\lambda_y) P_y(\lambda_y)$.

Now $|\bar{A}_x(\lambda_1)| \leq 1$, $|\bar{B}_y(\lambda_2)| \leq 1$ and it is easy to prove that <u>expectation values evaluated using (7) obey CHSH inequalities</u>. Namely:

$$s(\lambda_1,\lambda_2) = \bar{A}_x(\lambda_1)\bar{B}_y(\lambda_2) + \bar{A}_x(\lambda_1)\bar{B}_{y'}(\lambda_2) + \bar{A}_{x'}(\lambda_1)\bar{B}_y(\lambda_2) - \bar{A}_{x'}(\lambda_1)\bar{B}_{y'}(\lambda_2). \qquad (8)$$

$$|s(\lambda_1,\lambda_2)| \leq |\bar{A}_x(\lambda_1)||\bar{B}_y(\lambda_2)+\bar{B}_{y'}(\lambda_2)| + |\bar{A}_{x'}(\lambda_1)||\bar{B}_y(\lambda_2)-\bar{B}_{y'}(\lambda_2)| \leq 2. \qquad (9)$$

Using (5) we define $S = \sum_{\lambda_1,\lambda_2} s(\lambda_1,\lambda_2) P(\lambda_1,\lambda_2)$. Since $|S| \leq \sum_{\lambda_1,\lambda_2} |s(\lambda_1,\lambda_2)| P(\lambda_1,\lambda_2)$ from (9) we obtain immediately CHSH inequalities:

$$|S| = |E(A,B|x,y) + E(A,B|x,y') + E(A,B|x',y) - E(A,B|x',y')| \leq 2 \qquad (10)$$

A subtle point is that in SPCE one estimates expectation values using experimental protocol consistent with (3) not with (7). To estimate correlations using (7) one should repeat measurements (with the same setting) on the same correlated *pair of photons*, then average the results for each *pair* and finally average the results obtained for all the *pairs*. Such protocol is impossible to implement since you cannot repeat the measurements on the same *pair of photons* which was already absorbed by the detectors. CHSH can be proven because the summation in (7) over ($\lambda_x$, $\lambda_y$) destroys correlations created by a source. More detailed discussion of experimental protocols may be found in the next section and in [40].

## 4. KOLMOGOROV MODELS AND EXPERIMENTAL PROTOCOLS

Outcomes of any random experiment are described by a specific probability space Ω, σ-algebra F of all its sub-ensembles and a probabilistic measure μ. An ensemble $E \in F$ is called an event. A probability of observing the event E is given by $0 \leq \mu(E) \leq 1$. In statistics instead of Ω we use a sample space S which contains <u>only possible outcomes</u> of a studied random experiment. An extensive discussion of subtle notions of randomness and probability with applications to QM may be found in [44].

Kolmogorov probability theory similarly to quantum theory is *contextual* what means that different experiments are described by different probabilistic models using different spaces Ω. Please note that we use *contextual* in a specific probabilistic sense. For us probabilities are objective properties of random experiments performed in well-defined experimental contexts thus QM providing a probabilistic models for such experiments is a contextual theory.

Since in QM an act of observation is not a passive reading of pre-existing properties of physical systems thus models of a measurement process must introduce correctly the parameters describing measuring instruments.

There are few situations when we can use the same probability space and a joint probability distribution in order to describe different random experiments:

1. We have a set $O = (A_1, \ldots A_n)$ of compatible observables (properties) which may be observed or measured <u>in any order</u> on each member of some statistical population (group of people, identically prepared physical systems). In this case from a joint probability distribution of multivariate random variable we may deduce marginal probability distributions describing all random experiments in which only some subsets of the set $O$ are measured.

2. We have $n$ random experiments described by $n$ independent random variables (flipping of $n$ fair coins, rolling of $n$ fair dices) then, no matter how pairing of the experimental outcomes is done, outcomes are uncorrelated. A joint probability distribution is a product of probability distributions describing these $n$ independent random experiments.

Hidden variable models are not Kolmogorov models. In Kolmogorov models a sample space contains outcomes of real experiments. In hidden variable models parameters $\lambda$ are not outcomes of any feasible random experiment. These parameters are used to define invisible experimental protocols according to which observed outcomes of a random experiment might be obtained.

In Kolmogorov models a summation or integration over some variables, lead to marginal probability distributions describing realisable random experiments. In hidden variable models a summation or integration over some hidden variables defines, in general, new protocols describing different random experiments which cannot be performed. This is why, as we demonstrated in the previous section, the probabilistic hidden variable models (3) and (7) are not equivalent.

Nevertheless LRHV are isomorphic to Kolmogorov models describing the situation 1 and SHV to the models describing the situation 2. In contrast to the contextual probabilistic model (1-3) both of them assume hidden experimental protocols which cannot be implemented and which are inconsistent with the experimental protocols used in SPCE [40].

If hidden variables depend on the settings then Bell-type inequalities may not be proven [29, 34-40, 44-67]. Already in 1862, George Boole showed that whatever process generates a data set S of triples of variables $(S_1,S_2,S_3)$ where $S_i = \pm 1$, then the averages of products of pairs $S_iS_j$ in a data set S have to satisfy the equalities very similar to Bell inequalities [52, 55, 56]. In 1962 Vorob'ev [68] gave a concrete example showing that it was not possible to construct a joint probability distribution for any triple of pair-wise measurable dichotomous random variables.

Andrei Khrennikov [63, 64] constructed a generalised Kolmogorov probabilistic model describing SPCE experiment, with random choices of incompatible settings, and explained why in this model CHSH inequalities could not be proven.

As Theo Nieuwenhuizen said various hidden variable models used to prove Bell-type inequalities suffer from fatal theoretical *contextuality loophole* [66, 67] because they do not incorporate correctly supplementary parameters describing measuring instruments.

If Bell–type inequalities are violated by experimental data it only means that the models used to prove them are inappropriate for the experiments in which they are tested. This is why correlations violating Bell-type inequalities may be also found in different domains of science [61, 62].

## 5. ENTANGLEMENT AND INSTANTANEOUS INFLUENCES

Einstein, Podolsky and Rosen (EPR) [15] demonstrated that if two quantum particles interacted in the past and are separated then the outcomes of various measurements performed on them may be strongly correlated. According to QM measurement outcomes are produced in irreducibly random way thus if two physical systems are separated (free) the outcomes of measurements performed on them should not be correlated. Therefore separated particles seem to be entangled by some incomprehensible influences.

EPR concluded that this apparent paradox may be resolved if a statistical interpretation of the wave function is adopted. According to this interpretation the description of individual physical systems provided by QM is not complete. If outcomes of measurements, for example of linear momenta, performed on EPR particles are predetermined they are strongly correlated due to the conservation of total linear momentum of pair. This idea, incorporated in LHRV models failed to explain the correlations observed in SPCE. However as we discuss in section 8 spin projections are only defined and determined after the interaction with the measuring instrument in contrast to the values of linear momenta considered by EPR.

As we explained in section 2 QM does not predict perfect correlation or anti-correlation of spin projections [29, 34-40]. Therefore if Bob and Alice decide before the experiment to choose the same experimental setting when Alice obtains a click on one of her detectors, in a particular time-window, she does not know with certainty whether Bob registers a click on his detector at the same time window.

Therefore Alice's outcomes do not give any knowledge concerning Bob's outcomes. As we explain in section 7 for us the knowledge is not a degree of observer's belief quantified by some subjective probability. According to QM an outcome of a spin projection is only known when it is registered and as Peres [69] told: *unperformed experiments have no results*.

Another misconception is related to the description how entanglement is created. For example in his discussion of Delft group experiment Alain Aspect [70] writes: <u>*Mixing two photons*</u> *on a beam splitter and detecting them in coincidence* <u>*entangles*</u> *the electron spins on the remote NV centers*. This suggests an instantaneous influence of a single observed event on two remote physical systems. In our opinion the reason for correlations between faraway experiments may be a partial memory of the source carried by signals, like in (1), or particular preparations of far-away NV centers. <u>The observation of a particular coincidence signal does not create these particular preparations but gives only the information that such preparations occurred in remote places</u> [29].

Therefore to explain in intuitive way long range correlations observed in SPCE and predicted by QM we do not need evoke *quantum nonlocality* understood as a mysterious law of Nature according to which two distant random experiments produce perfectly correlated outcomes.

It is also claimed that to save locality of Nature one has to assume a s*uper-determinism* and deny the existence of free *will*. In the next section we show that such claims are unfounded.

## 6. BAYES THEOREM, FREE WILL AND MEASUREMENT INDEPENDENCE

According to model (1-3) experimentalists may use their *free will* to choose their experimental settings and explain imperfect correlations observed in SPCE in a locally causal way. We want to avoid here philosophical and/or theological discussion whether free *will* may exist or not. Our everyday decisions are conditioned by many uncontrollable factors but in physics our *freedom of experimenting* or *freedom of choice* is the unavoidable axiom.

In SPCE a question of existence of *free will* seems to be irrelevant since settings are selected using random number generators (RNG). It seems unnecessary and counter-productive to speculate whether RNG's and other measuring devices might be synchronized due to their shared past or not.

Therefore instead of talking about *free will* or *freedom of choice* one often prefers to talk about *measurement independence* defined as: *measurement settings can be chosen independently on any underlying variables describing the system* [71-73]. This assumption is expressed in terms of conditional probabilities: $P(x, y, \lambda) = P(x, y) P(\lambda)$, $P(x\, y|\, \lambda) = P(x, y)$ and

$$P(\lambda|\, x, y) = P(\lambda). \qquad (11)$$

If one omits to include parameters describing measuring instruments then it is reasonable to assume that a choice of remaining parameters and a choice of the settings are independent events in the usual probabilistic sense. This is why in SHV (4) and in LRHV (5, 6) probability distributions of hidden variables do not depend on the setting (x, y). As already Bell [1] noticed if $P(\lambda|x, y) \neq P(\lambda)$ then Bell-type inequalities cannot be proven.

If $P(\lambda|x, y) \neq P(\lambda)$ then using Bayes Theorem one may prove that $P(x, y|\lambda) \neq P(x, y)$ what is wrongly interpreted as: *the experimentalists have no freedom of choosing their experimental settings.*

As we mentioned already in section 2 such interpretation is based on incorrect interpretation of conditional probabilities. We will give below a detailed explanation.

In Kolmogorov probabilistic model Bayes Theorem is simply a definition of conditional probabilities [44]. Let us consider three events $E_1$, $E_2$ and A (in mathematical statistics and in physics by events we understand a subset of outcomes of some random experiment) such that $E_1 \cap E_2 = \emptyset$, $E_1 \cap A \neq \emptyset$ and $E_2 \cap A \neq \emptyset$ where $\emptyset$ denotes an empty set. Conditional probabilities $P(A|E_i)$ and $P(E_i|A)$ are simply defined by a formula:

$$P(E_i \cap A) = P(E_i|A) P(A) = P(A|E_i) P(E_i) \tag{12}$$

Since events $E_1$ and $E_2$ are mutually exclusive we may write also a total probability formula:

$$P(A) = P(A|E_1) P(E_1) + P(A|E_2) P(E_2) \tag{13}$$

Conditional probabilities, in general, say nothing about causal relation between the events A and $E_i$. Let us consider two random experiments:

1. In experiment 1 we have in a box red and white balls which are: big or small. A random experiment consists in drawing a ball with replacement and registering the result. A sample space S={(r, b), (r, s), (w, b), (w, s)} and let us assume that P(r, b)=1/8, P(r, s)=1/4, P(w, b)=3/8 and P(w, s)=1/4. We define events $E_1$={(r, b), (r, s)}, $E_2${(w, b), (w, s)} and A=={(r, b), (w, b)} and calculate conditional probabilities: P(A|$E_1$)= (1/8)/ (1/8 +1/4)=1/3 and P($E_1$| A)= (1/8)/ (1/8 + 3/8) =1/4 . The meaning of these conditional probabilities is: if we repeat our experiment: "1 out of 3 drawn red balls is likely to be big" and "1 out of 4 drawn big balls is likely to be red". There is no causal relation between $E_1$ and A.

2. Experiment 2 is a randomized drug trial in which a selected group of people with a moderately high blood pressure takes daily a dose of a candesartan (C) or a placebo (P). After 1 month the blood pressure is measured several times to check whether targeted lower levels were achieved: yes (Y) or not (N). Corresponding events are: $E_1$={(C,Y), (C,N)}, $E_2$= ={(P,Y), (P,N)} and A=={(C,Y), (P,Y)}. There is probably a *causal relation* between "taking a candesartan" and "lowering of a blood pressure" which is quantified by a probability P (A|$E_1$). There is no causal interpretation attached to the conditional probability P ($E_1$|A) which tells us only "how likely an individual on a study whose blood pressure decreased was taking a candesartan".

The experiment 2 is a good example of a standard application of Bayes Theorem and of Bayesian reasoning.

In Kolmogorov approach any probability assigned to a specific event must in principle be liable to verification. In hidden variable models $P(x, y, \lambda)$ are inaccessible to any verification. Nevertheless we may still try to arrive to some conclusions using Bayes Theorem and assuming that $P(x, y, \lambda) = P(\lambda | x, y)) P(x, y)$. Let us see it using our model (1).

$$P((\lambda | x, y) = P(\lambda_1, \lambda_2) P_x(\lambda_x) P_y(\lambda_y) P(x,y)/P(x,y) = P(\lambda_1, \lambda_2).P_x(\lambda_x) P_y(\lambda_y) \qquad (14)$$

$$P((x, y| \lambda) = P(\lambda_1, \lambda_2) P_x(\lambda_x) P_y(\lambda_y) P(x,y)/ P(\lambda_1, \lambda_2) P_x(\lambda_x) P_y(\lambda_y) P(x,y) = 1 \qquad (15)$$

To prove (15) we use the identity $P(\lambda_x, \lambda_y) = P_x(\lambda_x) P_y(\lambda_y) P(x, y)$. The meaning of (15) is the following: *if an "event" ($\lambda_x, \lambda_y$) occurs then the settings (x, y) were used*. It does not mean that ($\lambda_x, \lambda_y$) have any *causal influence* on the choice of the settings. Therefore all claims [21, 31] that Bayes Theorem proves the opposite are simply incorrect.

People often do not distinguish between colloquial *independence* and *stochastic independence of events*. They forget that if $E_i \cap A = E_i$ (events are dependent) then *$P(A | E_i) = 1$ but it does not mean that there is any causal influence of the event $E_i$ on the event A.*

Since the *setting dependence of hidden variables* does not constrain experimenters' *freedom of choice* thus it should be called *contextuality* instead of being called *violation of measurement independence*.

Michael Hall [72, 73] arrived recently, using a different reasoning, to a similar conclusion: *a violation of measurement independence is not automatically inconsistent with apparent experimental freedom*.

Bayesian reasoning is a powerful tool in modern mathematical statistics having many applications in various domains of science including physics. This is why some physicists were tempted to interpret quantum probabilities as subjective judgements of human agents and claimed that such interpretation called *Qbism* (*Quantum Bayesianism*) is the best interpretation of QM free of paradoxes. We disagree with this claim and we explain it in the next section

## 7. SUBJECTIVE PROBABILITIES, QBISM AND PHYSICAL REALITY

According to QBism: quantum probabilities are probability assignments made by human agents [74-76]. For QBists entire *purpose of quantum mechanics* and physics in general "*is to enable any single agent to organize her own degrees of belief about the contents of her own personal experience*" [75]. This point of view is has been criticised [77-79].

In our opinion the aim of physics is to discover laws of Nature governing objectively existing external world. Physicists are not verifying their personal beliefs about their observations but search for a mathematical abstract description allowing to explain and to predict in a quantitative way physical phenomena observed and those to be discovered. These laws do not depend on the existence and on the presence of human agents.

Of course we perceive the surrounding world by our senses and we probe it using experimental devices constructed by us. If our senses and brains were different we would probably grasp different aspects of this world if we were able to do it.

It is amazing that the laws of Nature, we were able to discover and the theories we constructed, allow us to produce new materials, cure diseases and explore the Solar system and beyond. Perhaps our success is due to the fact that our senses and brains are the product of the evolution which had been governed by the laws of Nature which we were able only to discover in last centuries.

These laws of Nature governed physical phenomena long before Homo sapiens started their quest for the explanation. Stars were created and vanished, tectonic plates were moving, planets were orbiting around the Sun, seasons and tides were changing periodically. We can continue with examples from the biology: animal migration patterns were repeating courtship rituals and mating were coded in genes; the evolution of species was driven by the environmental changes etc.

In quantum phenomena invisible signals and particles prepared by some sources interact with macroscopic devices and after a substantial magnification produce clicks on various detectors. Quantum theory provides an abstract mathematical model allowing: to make objective quantitative predictions about the energy levels of atoms and molecules, to make probabilistic predictions concerning a statistical scatter of outcomes registered by various macroscopic instruments etc.

Quantum probabilistic models contain often free parameters which can be treated as Bayesian priors. The best estimates of the unknown values of these parameters are obtained using the maximum likelihood model based on Bayesian approach.

This does not allow interpreting wave functions (state vectors) as mathematical entities corresponding to subjective beliefs of human agents. Two high energy protons from cosmic rays, when they collide far away from human agents, are as likely to produce several mesons $\pi$ as if they collided on the Earth. This is only true if the branching ratios for different reaction channels do not depend on the presence of gravitation etc.

The discussion of *quantum nonlocality* and statements ''quantum correlations are necessarily between time-like events'' [75] do not explain why strong correlations between space-like events in SPCE may and do exist.

In SPCE a source is sending two signals to far away detectors, clicks are produced and registered. After gathering several clicks correlations between these clicks can be estimated and these estimations outputted by on-line computers. The existence of correlated outcomes does not depend whether some human agents perceived these clicks or not.

QBists forget that quantum mechanics and quantum field theory are much more than quantum information and that physicists are not interested in subjective experiences of various agents but they are trying to deduce the laws of Nature governing the external world.

A profound discussion of human efforts to understand and to model *Physical Reality* may be found in a stimulating book of Hermann Weyl [80].

## 8. THE BOHR-EINSTEIN QUANTUM DEBATE

In his viewpoint Alain Aspect [70] resumes discussions and experiments performed to check Bell inequalities. He concludes that Einstein was wrong and that recent experiments [12-14] *closed the door on Einstein and Bohr's quantum debate*.

We disagree with this conclusion since as we explained in preceding sections the violation of Bell-type inequalities proves only that outcomes in SPCE are neither predetermined nor produced in irreducibly random way. The hidden variable models LRHV and SHV suffer from *contextuality loophole* since they do not incorporate correctly supplementary parameters describing measuring instruments [66, 67]. If such parameters are introduced Bell-type inequalities cannot be proven and correlations may be explained in causally local way.

The Bohr–Einstein quantum debate is more fundamental. *Does QM provide a complete description of individual physical systems and of Physical Reality and in what sense?*

The failure of SHV to explain SPCE seems to confirm Einstein's intuition that quantum probabilities are reducible and that QM may be emergent from some underlying theory giving more detailed locally causal description of phenomena.

According to Bohr a satisfactory description of Physical Reality is given in terms of indivisible complementary quantum phenomena produced when identically prepared physical systems are analysed by various macroscopic experimental instruments. For Bohr quantum probabilities are irreducible and a more detailed locally causal sub-quantum description of observed phenomena is impossible.

For Einstein wave functions and density matrices describe only ensembles of identically prepared physical systems. The wave function reduction is not instantaneous and a reduced wave function provides a description of a new ensemble of physical systems. Einstein objected to Bohr's claim that probabilistic predictions of QT provided the most complete description of individual physical systems and believed that more detailed description of atomic objects is possible [16,17, 33, 36, 37]. In particular he said to Pauli: *like the moon has a definite position* *whether or not we look at the moon, the same must hold for the atomic objects, as there is no sharp distinction possible between these and macroscopic objects.*

Let us note that Einstein is talking about positions, when nobody observes them, and not about predetermined outcomes of position measurements. There is a fundamental difference between a position and a spin projection. Spin projection is a *contextual property* which is created in the interaction of electrons and photons with Stern-Gerlach instruments or with polarization beam splitters (PBS) respectively. A position does not need to be observed in order to be believed to exist.

For example let us consider an electron. Many physicists, not listening to Bohr, *imagine* an electron as a point-like particle surrounded by the electromagnetic field it creates. A point-like particle by definition must have unknown but precise position in some reference frame.

We know the electron's rest mass and its electric charge no matter how we measure it. When passing through a bubble chamber the electron ionizes atoms creating a *macroscopic picture* of electron's trajectory. We can neither measure sharply its position nor linear momentum. However we can prepare in accelerators, using classical relativistic electrodynamics beams of the electrons, protons, heavy ions etc., with a little spread in their linear momentum and energy, and we may project these beams on different targets to study their collisions.

In contrast a spin projection is not electron's predetermined attribute. The experiments [6-14] confirm this particular character of spin projections but they have nothing to say about masses, charges, momenta etc. In contrast to electron no intuitive mental image of photon may be created.

It is true that space-time localisation loses its operational foundation in the micro-world and any attempt to strictly localize an elementary particle would lead probably to its destruction and/or to production of many other particles.

This is why to study interactions of elementary particles one uses abstract state vectors described only by linear momenta, energy, spin and several internal quantum numbers. Feynman graphs are mnemonic tools helping to calculate, in the perturbative expansion, various contributions to transition amplitudes. Nevertheless physicists imagine electrons and quarks as point-like particles surrounded by a cloud of virtual photons, gluons etc. The interactions between these elementary constituents of matter are described as exchanges of virtual photons and gluons. Moreover the "radius" of proton, ranges of various interactions and lifetimes of many unstable particles and resonances are estimated. Thus we see that in particle physics a space-time description of invisible events is not abandoned.

Human beings in order to "understand", they have to create mental *images* and such images cannot be created without a notion of a space-time. Einstein would probably agree with Bohr that Feynman graphs are not faithful images of what really is happening. Nevertheless strangely enough such *mental images* inspire physicists to improve their mathematical models what leads to discovery of new phenomena, new elementary particles and resonances.

If physicists listened to Bohr (with whom we I agree in many points) and abandoned constructing *mental images* a progress of science would be compromised.

This is why recent experiments of Yves Couder and collaborators [81], with bouncing droplets (http://dualwalkers.com) may help perhaps to develop one day more intuitive understanding of a wave–particle duality and of quantum phenomena.

There is still no general agreement how to interpret QM, QFT and how to reconcile them with the general relativity. In our opinion a rigorous epistemological and ontological analysis of quantum field theory is still lacking.

<u>Incorrect interpretation of QT and/or incorrect *mental images* of quantum particles and measurements   lead to paradoxes and to speculations which are a pure science fiction.</u>

Leslie Ballentine said in his book: "*Once acquired, the habit of considering an individual particle to have its own wave function is hard to beak. Even though it has been demonstrated strictly incorrect*" [33].  Statistical contextual interpretation, inspired by Bohr and Einstein, seems to be the most cautious [29, 33, 35-37, 61]. Several arguments in favour of statistical interpretation of QT and how it may account for all properties of quantum measurements, including the uniqueness of individual outcomes were given by Allahverdyan et al. [82, 83].

According to statistical interpretation quantum states and operators are treated as mathematical abstract tools enabling quantitative predictions to be compared with the experimental data.  Whether QM is emergent from some more detailed theory of quantum phenomena is an open question.

Since SHV failed to explain the correlations observed in SPCE, quantum probabilities are not irreducible, as Bohr claimed, If they are not irreducible they may emerge as in (1) from some more detailed sub-quantum description what was suggested by Einstein.  However, as Bohr insisted, quantum phenomena are contextual therefore any subquantum model of quantum measurement has to be contextual (include a description of measuring instruments).

Several attempts are made to provide more detailed description of various quantum phenomena see for example [84-86].

<u>We  don't  even  know whether quantum theory  is  *predictably complete* ?</u>   In order to answer this question a more detailed analysis of experimental data is  needed  [36, 37, 87-89] .

In view of all these arguments contrary to Aspect's claim the results of experiments [6-14] do not show that Einstein was wrong and Bohr was right. In our opinion they show that Einstein's and Bohr's ideas may be in some sense reconciled and that *<u>the door on Einstein and Bohr's quantum debate</u> <u>cannot be closed</u>*

# 9. CONCLUSIONS

The Bohr-Einstein quantum debate has been inspiring physicists for last 80 years. Does quantum theory provide a complete description of individual physical systems? Is Nature local or not local? How can we reconcile apparent indeterminism of quantum phenomena with apparent locality and causality of macroscopic world we are living in?

Efforts to answer these and other questions not only allowed incredible technological progress but also: computers, smart-phones and internet changed our social relations and culture. We are living now a second quantum revolution this is why it so important to find the explanation of quantum phenomena without evoking quantum magic. Magical explanations are usually counter-productive and misleading.

The violation of Bell-type inequalities demonstrated only that local realistic and stochastic hidden variable models failed to describe correlations observed in SPCE and that measured values of spin projections <u>are neither predetermined nor produced in intrinsically random way</u>.

Probabilistic models LRHV and SHV did not incorporate correctly supplementary parameters describing measuring instruments. If setting dependent parameters are introduced correlations observed in SPCE can be explained in causally local way and Bell-type inequalities may not be proven. As we demonstrated in the section 6, contrary to a general belief, the introduction of such setting-dependent parameters is not in conflict with *experimenters' freedom of choice*.

In recent tests of *local realism* a considerable effort was made to close the locality *loophole* and to create experimental conditions which made impossible causal influences between remote experimental devices. This constraint reduced significantly the size of post-selected samples.

As we demonstrated such <u>influences are not necessary for the existence of strong correlations</u> between outcomes of distant random experiments thus the correlations observed should not depend on how the settings are changed. It would be interesting to check, in the same experiment, whether closing or not closing the locality-loophole leads to <u>statistically significant differences</u> between estimated correlations or not. We claim that it should not.

Then instead of testing various Bell-type inequalities one may concentrate on testing QM predictions for different angles by creating larger samples for each setting. One has to use synchronized time-windows and select those in which both clicks were observed. One should not only estimate the correlations but also probability distributions of single clicks. It would allow to test homogeneity [29, 90, 91] of experimental samples, for each choice of settings, and to test whether some anomalies found earlier in Aspect's and Weihs' data would reappear [92-94].

John Bell insisted that the question of completeness of QM cannot be answered using a mathematical or philosophical reasoning but only by experiments. The experiments [4-16] eliminated LRHV and SHV but were not able to give a definite answer. As Bell said in a different context: "*What is proved by impossibility proofs is lack of imagination*".

If QM is an emergent theory, as claimed by Einstein, then experimental time series may contain fine structures which are not predicted by QM and which might have been averaged out in a standard statistical analysis of experimental data. Thus to check whether QT is predictably complete one has to study experimental time series more in detail then it is usually done [87-89].

The violation of various Bell-type inequalities does not allow to conclude that <u>*Nature is nonlocal and that God plays dice*</u>. Therefore Bohr-Einstein quantum debate may not be closed. The continuation of this debate is not only important for a better understanding of Nature but also for various practical applications of quantum phenomena.